# Observation of Active Sites for Oxygen Reduction Reaction on Nitrogen-doped Multilayer Graphene

*Tan Xing,[†] Yao Zheng,[‡] Lu Hua Li,[†]\* Bruce C.C. Cowie,[§] Daniel Gunzelmann,[†] Shi Zhang Qiao,[‡] Shaoming Huang[⊥] and Ying Chen[†]*

[†]Institute for Frontier Materials, Deakin University, Geelong Waurn Ponds Campus, VIC 3216, Australia

[‡]School of Chemical Engineering, The University of Adelaide, Adelaide, SA 5005, Australia

[§]Australian Synchrotron, 800 Blackburn Road, Clayton, VIC 3168, Australia

[⊥]Nanomaterials and Chemistry Key Laboratory, Wenzhou University, Wenzhou 325027, China

ABSTRACT

Active sites and catalytic mechanism of nitrogen-doped graphene in oxygen reduction reaction (ORR) have been extensively studied but are still inconclusive, partly due to the lack of an experimental method that can detect the active sites. It is proposed in this report that the active sites on nitrogen-doped graphene can be determined *via* the examination of its chemical composition change before and after ORR. Synchrotron-based X-ray photoelectron spectroscopy analyses of three nitrogen-doped multilayer graphene samples reveal that oxygen reduction intermediate OH(ads) which should chemically attach to the active sites remains on the carbon atoms neighboring pyridinic nitrogen after ORR. In addition, a high amount of the OH(ads) attachment after ORR corresponds to a high catalytic efficiency and vice versa. These pinpoint that the carbon atoms close to pyridinic nitrogen are the main active sites among the different nitrogen doping configurations.

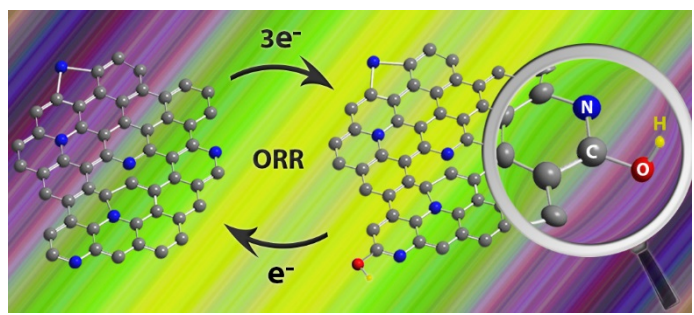

Fuel cells are a promising sustainable and renewable energy source. The practical use of fuel cells highly depends on the development of suitable catalysts for oxygen reduction reactions (ORR). Nitrogen-doped graphene is a potential carbon-based ORR catalyst of low cost, high stability and high efficiency.[1-4] In spite of extensive studies, the exact catalytic mechanism of nitrogen-doped graphene is still under debate. A fundamental disagreement lies in how nitrogen contributes to the catalysis. To answer the question, the key is to determine the contribution of three nitrogen doping configurations, namely pyridinic, graphitic and pyrrolic, to the catalytic performance, which is still inconclusive. Some studies support that pyridinic nitrogen is catalytically active;[5-10] while others suggest graphitic nitrogen is effective.[11-17] There are also claims that both pyridinic and graphitic nitrogen contribute to the catalytic property but with different roles.[18,19] Till now all the experimental determinations of the active sites on nitrogen-doped graphene are indirect. The controversial results from the previous experiments suggest that there is a need to develop a method that is able to observe the active sites and elucidate the catalytic mechanism. It was demonstrated before that intermediates of a catalytic reaction could be *ex situ* analysed on the surface of the catalyst.[20,21] Here, we report that chemically adsorbed oxygen reduction intermediates can be detected on the nitrogen-doped multilayer graphene after ORR using X-ray photoelectron spectroscopy (XPS). This not only implies that nitrogen close to carbon in nitrogen-doped graphene is involved in the catalysis of ORR process and, more importantly, can be used to observe the active nitrogen sites and pinpoint the ORR pathway.

RESULTS AND DISCUSSION

Three multilayer graphene samples were prepared from graphene oxide using different nitrogen sources and doping methods for different nitrogen doping configurations and concentrations: treatment by ammonia hydroxide, heating under ammonia gas and reaction with melamine (see Experimental Section). For simplicity, the three samples are named $G\text{-}NH_3 \cdot H_2O$, $G\text{-}NH_3$ and $G\text{-}C_3N_4$, corresponding to the nitrogen doping methods. Typical transmission electron microscopy (TEM) images of the nitrogen-doped multilayer graphene nanosheets which are a few nanometers thick are shown in Figure 1.

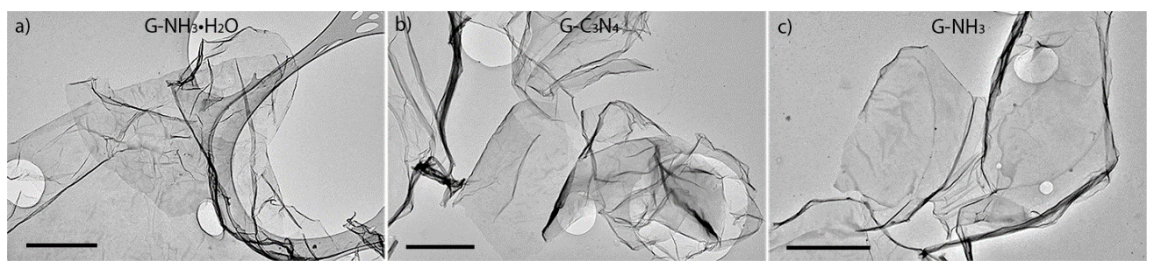

Figure 1. TEM images of a) $G\text{-}NH_3 \cdot H_2O$, b) $G\text{-}C_3N_4$ and c) $G\text{-}NH_3$ nitrogen-doped multilayer graphene samples. All scale bars are 2 µm.

Synchrotron-based XPS of high resolution and high sensitivity was used to investigate the chemical composition change on the three nitrogen-doped multilayer graphene samples before and after ORR. Before ORR, the nitrogen contents in G-NH$_3$·H$_2$O, G-C$_3$N$_4$ and G-NH$_3$ are 6.0%, 5.7% and 6.8%, respectively. The least-square fittings of the XPS spectra in the nitrogen 1s region before ORR are displayed in Figure 2 (upper row). It can be seen that G-NH$_3$·H$_2$O and G-NH$_3$ samples have relatively higher contents of pyrrolic nitrogen at 399.8 eV (red), slightly less pyridinic nitrogen at 398.5 eV (blue), and much less graphitic nitrogen at 401.2 eV (purple). G-C$_3$N$_4$, on the other hand, contains more pyridinic nitrogen and less but comparable amounts of pyrrolic and graphitic nitrogen. After ORR, the three samples show different changes of pyridinic, pyrrolic and graphitic nitrogen contents (lower row in Figure 2). For G-NH$_3$·H$_2$O, the pyridinic nitrogen peak decreases while the "pyrrolic" peak increases; for G-C$_3$N$_4$, the graphitic content increases and the pyridinic and "pyrrolic" contents are steady; G-NH$_3$ shows a depressed "pyrrolic" peak. As shown below, these changes correspond to very different catalytic behaviors and functions of the different nitrogen configurations during the ORR process.

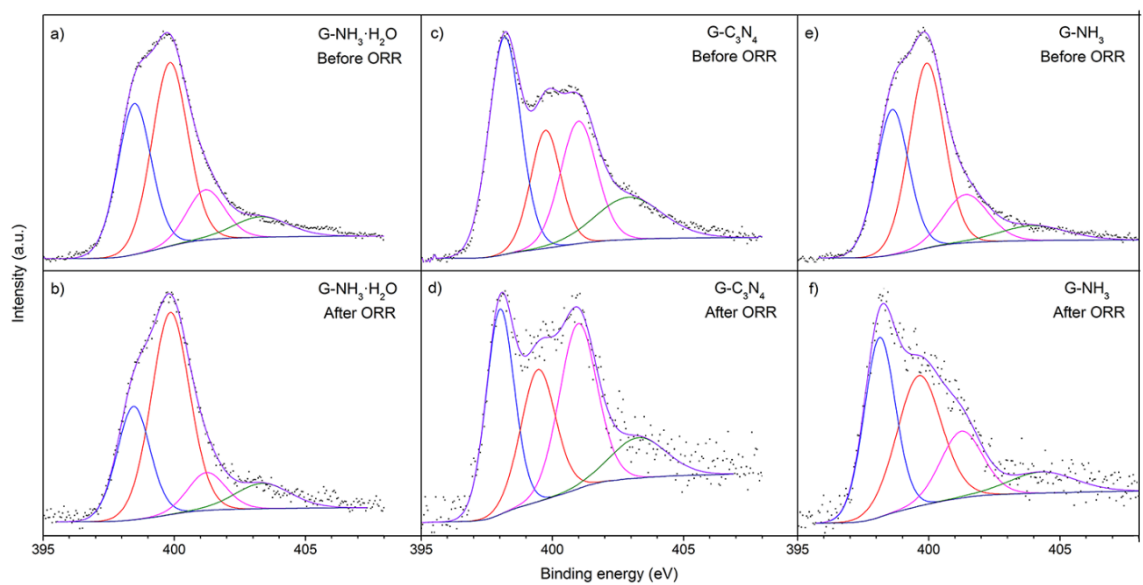

Figure 2. Nitrogen 1s XPS spectra of the three multilayer graphene samples before and after ORR. The least-square fitted peaks are pyridinic nitrogen at 398.5 eV (blue), "pyrrolic" nitrogen at 399.8 eV (red), graphitic nitrogen at 401.2 eV (purple) and nitrogen oxide at 403 eV (green).

The observed changes in nitrogen contents before and after ORR cannot be attributed solely to nitrogen loss. Nitrogen may be lost *via* a nitrogen oxidation process during ORR, which could change the contents of the different nitrogen configurations. Generally, pyrrolic nitrogen is less stable than pyridinic and graphitic nitrogen and more prone to the oxidation and loss during ORR process.[22,23] In other words, the pyrrolic nitrogen peak should relatively decrease after ORR. However, G-NH$_3$·H$_2$O shows the opposite trend with an increased "pyrrolic" peak after ORR.

For clues to the reason for the relative change of the nitrogen contents, we can turn to the carbon and oxygen XPS results. Both the oxygen and carbon 1s XPS spectra reveal that the content of –OH, one of the intermediate products of ORR, increases on G-NH$_3$·H$_2$O after ORR. Figure 3 compares the oxygen 1s XPS spectra of G-NH$_3$·H$_2$O before and after ORR. It can be seen that after ORR the C(aliphatic)–OH/C(aliphatic)–O–C(aliphatic) peak at 532.0 eV (red) relatively decreases in intensity and the C(aromatic)–OH at 533.3 eV (blue) increases in intensity.[24,25] This suggests that compared to the other oxygen-containing groups, the relative content of –OH group attached to aromatic carbon rises after ORR. The carbon 1s XPS spectra also show that the C–OH peak at 285.4 eV[24] (purple) dramatically strengthens in G-NH$_3$·H$_2$O after ORR (Figure 4).

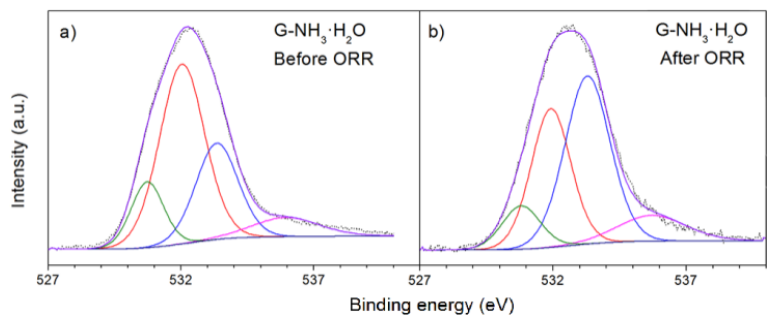

Figure 3. Oxygen 1s XPS spectra of G-NH$_3$·H$_2$O a) before and b) after ORR. The fitted peaks are C=O at 530.8 eV (green), C(aliphatic)–OH/C(aliphatic)–O–C(aliphatic) at 532.0 eV (red), C(aromatic)–OH at 533.3 eV (blue) and chemisorbed water molecules at 535.7 eV (purple).

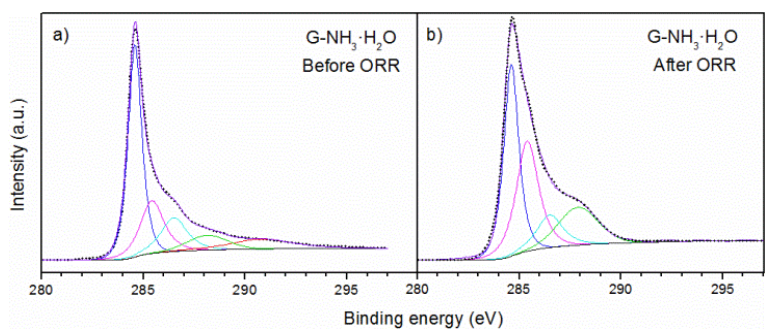

Figure 4. Carbon 1s XPS spectra of G-NH$_3$·H$_2$O a) before and b) after ORR. The fitted peaks are C–C/C=C at 284.6 eV (blue), C–OH at 285.4 eV (purple), C–O–C at 286.5 eV (cyan), C=O at 288.1 eV (green) and –COOH at 290.8 eV (red). The contribution from Nafion to the carbon 1s XPS spectrum has been subtracted (see Supporting Information).

Based on these evidence, the changes of the nitrogen 1s XPS profiles in Figure 2 could be caused by the chemical environment change of the doped nitrogen during ORR, i.e. the attachment of –OH to the carbon atoms bond to the nitrogen. It has been reported that the attachment of –OH to pyridone can cause an upshift of the nitrogen binding energy from 398.8 to 400.2 eV in XPS.[22,26,27] Similarly, when an –OH attaches to the carbon neighboring pyridinic nitrogen in graphene, the nitrogen binding energy should shift upwards to an energy very close to the 399.8 eV position of the pyrrolic peak. Therefore, the observed intensity decrease of the pyridinic peak and intensity increase of the "pyrrolic" peak in G-NH$_3$·H$_2$O after ORR could be caused by the change of the pristine pyridinic nitrogen to the pyridinic nitrogen with a neighboring carbon attached to –OH, as illustrated in Figure 5. In other words, the increase of the "pyrrolic" nitrogen after ORR in XPS is not caused by the increase of pyrrolic content, but the increased content of the –OH attached pyridinic nitrogen which has a very similar XPS peak position to the pyrrolic nitrogen. The existence of the –OH attached pyridinic nitrogen is also supported by the increased full width at half maximum (FWHM) of the least-square fitted "pyrrolic" peak at 399.8 eV from 1.61 eV before ORR to 1.71 eV after ORR, suggesting the existence of the –OH attached pyridinic nitrogen peak in the region. The unchanged and decreased intensities of the 399.8 eV XPS peak for G-$C_3N_4$ and G-$NH_3$ suggest that much less –OH attachments are formed to the carbon atoms neighboring pyridinic nitrogen in the two graphene samples and pyrrolic nitrogen is slightly lost during ORR.

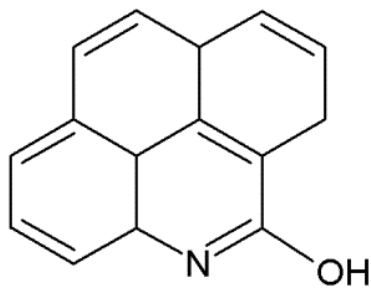

Figure 5. A diagram of the chemical structure of –OH attached to the carbon neighboring pyridinic nitrogen, leading to a bonding energy upshift of the pyridinic nitrogen in XPS.

The above interpretation on the changed nitrogen contents observed in XPS is in excellent agreement with Fourier transform infrared spectroscopy (FTIR) analyses. The contents of –OH on the three samples before and after ORR have been measured by FTIR.[28,29] In Figure 6, the characteristic –OH peak centered at ~3440 cm$^{-1}$ becomes much stronger for G-$NH_3$·$H_2O$ after ORR; in contrast, G-$C_3N_4$ and G-$NH_3$ have eithor unchanged or decreased –OH contents after ORR. It should be mentioned that the –OH signals in FTIR are not due to water, because all the samples were heated at 65 °C in vacuum for 24 h. Additionally, solid-state nuclear magnetic resonance (NMR) spectroscopy shows that most hydrogen is in the form of C–OH in G-$NH_3 \cdot H_2O$ after ORR and no $H_2O$ signal was detected (see Supporting Information, Figure S2).[30,31]

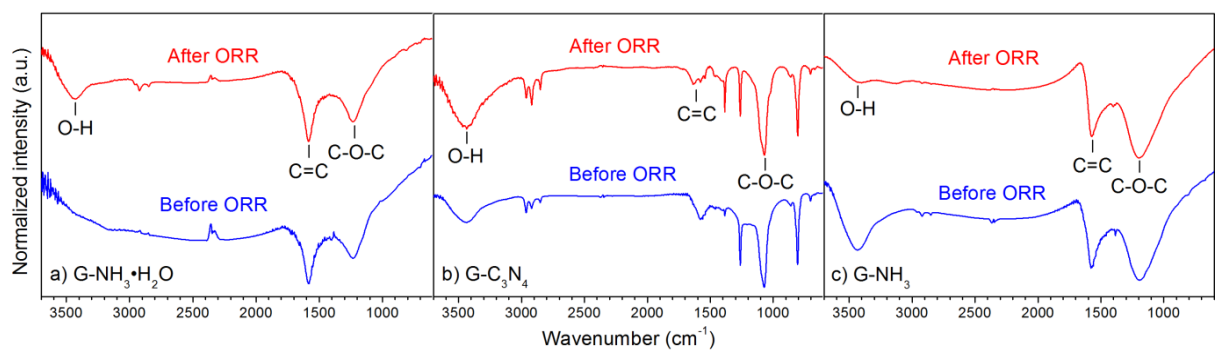

Figure 6. FTIR spectra of a) G-$NH_3 \cdot H_2O$, b) G-$C_3N_4$ and c) G-$NH_3$ before (blue) and after (red) ORR. For G-$C_3N_4$, the peak at 800 cm$^{-1}$ may be C−N related and the one at 1260 cm$^{-1}$ may be due to C–N or C−H.[32]

Intriguingly, G-$NH_3 \cdot H_2O$ which experiences the highest increase of the –OH attachment to pyridinic nitrogen after ORR shows the best catalytic performance; G-$C_3N_4$ and G-$NH_3$ which have no increase or even decrease of the –OH attachment after ORR show much worse ORR properties. The ORR catalytic performances of the three samples in alkaline electrolyte are compared in Figure 7 (the current-potential curves can be found in the Supporting Information, Figure S3). For G-$NH_3 \cdot H_2O$, the onset potential of ORR is about −0.13 V and the electron transfer number is 3.75 with a dynamic limiting current ($J_k$) of about 15 mAcm$^{-2}$ at 0.6 V vs Ag|AgCl. Both G-$C_3N_4$ and G-$NH_3$ show more negative onset potentials, lower electron transfer numbers and $J_k$. The correlation between the increased content of the –OH attached pyridinic nitrogen and the ORR performance can be used to determine the active sites for ORR on nitrogen-doped graphene.

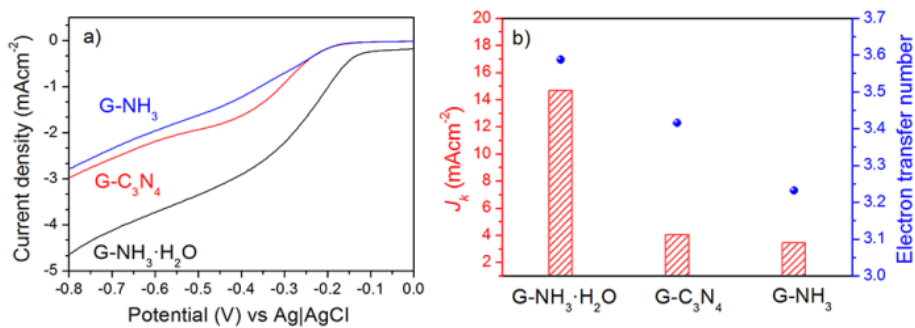

Figure 7. The catalytic performance of the three nitrogen-doped multilayer graphene samples ($J_k$ and electron transfer number are calculated at −0.6 V vs Ag|AgCl). The electron transfer number and the kinetic limiting current are determined from slopes of Koutecky-Levich plots (see Supporting Information).

In the commonly accepted four-electron associative ORR mechanism of nitrogen-doped graphene in alkaline solution, intermediate OH(ads) is formed from O(ads) with the addition of $H_2O$ and one electron.[3,33] It is predicted that the OH(ads) is attached to the catalytic core or active site *via* a chemical bond.[9,33,34] Therefore, the ORR active sites can be determined from the presence of intermediate OH(ads). Our observation of the −OH attached pyridinic nitrogen after the electrochemical reaction suggests that pyridinic nitrogen plays an important role in the ORR process and the nearby carbon should be the main active sites, in good agreement with some of the previous theoretical and experimental studies.[5-10,35,36] According to these studies, the neighbor carbon atoms of pyridinic nitrogen have favored atomic charges and spin density distributions that can induce the ORR process: the absorption of the intermediate products, the formation of C–O bond and the disassociation of O–O bond.[9,34] The carbon bond to graphitic nitrogen should not be the main active site in the three tested samples, because according to XPS, the content change of graphitic nitrogen before and after ORR is much less significant and therefore presents no strong relation with the catalytic performance.

Another very interesting point from this study is that the dominant type of nitrogen doping configuration may not necessarily correlate with the ORR performance of nitrogen-doped graphene. Before ORR, G-NH$_3$·H$_2$O and G-NH$_3$ have similar contents of nitrogen and oxygen (about 13% in G-NH$_3$·H$_2$O and 14% in G-NH$_3$). The contents of pyridinic, pyrrolic and graphitic nitrogen in the two samples are similar too (Figure 2a and e). In addition, the three samples show similar Raman spectra and electrochemical impedance values, indicating their similar crystallinity (defect levels) and electrical conductivity (see Supporting Information, Figure S4 and S5). The electrochemical impedance tests also imply comparable wettability of the three samples, consistent to their capability to be dispersed in water. However, the two samples show dramatically different catalytic properties (Figure 7). This clearly shows that the pyridinic or graphitic nitrogen content before ORR cannot be used to predict or explain ORR performance. This also raises a question: why do the two samples with similar pyridinic contents show dramatically different ORR performance if the pyridinic sites are the main catalytic active sites? One presumption is that pyridinic nitrogen in graphene may have different ORR properties. It has been demonstrated by theoretical calculations that the microstructure of nitrogen doping could lead to different ORR activities.[16,37,38] So, although the total amounts of pyridinic nitrogen are similar in G-NH$_3$·H$_2$O and G-NH$_3$, it is likely that the pyridinic nitrogen in G-NH$_3$·H$_2$O and G-NH$_3$ have different dominant microstructures. For example, our oxygen 1s XPS spectra in Figure 3 show that the –OH is mainly attached to aromatic carbon rather than aliphatic carbon in the ORR, suggesting that the pyridinic nitrogen on the edge should be more likely to involve in the catalytic process. So it is possible that G-NH$_3$·H$_2$O has more pyridinic nitrogen on the edge than G-NH$_3$. According to theoretical calculations,[37] other microstructures, such as the edge type (zig-zag or arm-chair) and neighboring nitrogen doping, may also affect the ORR activity of pyridinic nitrogen. Unfortunately, the microstructure of pyridinic nitrogen in graphene is very difficult to determine. Nevertheless, as demonstrated by this study, the examination of nitrogen-doped graphene after ORR can reveal the amount of active sites *via* the appearance of oxygen reduction intermediate, i.e. OH(ads). The ammonia hydroxide doping gives the best ORR performance among the three nitrogen doping strategies, but the mechanism for forming different microstructures of pyridinic nitrogen requires further investigation. This could be important for tailoring nitrogen doping site and further improving the ORR efficiency of nitrogen-doped graphene.

## CONCLUSION

In summary, the examination of chemical composition change of nitrogen-doped multilayer graphene before and after ORR can be used to detect the chemically attached ORR intermediate i.e. OH(ads) and determine the catalytic active sites. The good correlation between the amounts of the OH(ads) attachment and the ORR performances from the three nitrogen-doped multilayer graphene samples suggests that the carbon atom neighboring pyridinic nitrogen plays an important role in the ORR process and should be the main active sites in the tested samples. This study also implies that the location or microstructure of pyridinic nitrogen can affect its catalytic efficiency.

## EXPERIMENTAL SECTION

Graphene oxide (GO) was prepared from graphite flakes (Sigma, 325 mesh) using a slightly modified Hummers' method.[39,40] The GO powder was collected by lyophilization. Three types of nitrogen-doped multilayer graphene were synthesized *via* different nitrogen sources and incorporation methods. For G-NH$_3$·H$_2$O, the initial GO solution (~1 mg/mL) was mixed with 5 mL 25% NH$_3$·H$_2$O, then the mixture was stirred and heated at 90 °C for 12 h to complete nitrogen incorporation. Resultant solution was then washed by water to remove base and lyophilized to get the final product in a dry powder form. G-NH$_3$ was prepared by annealing the GO powder in 20% NH$_3$/Ar at 600 °C for 5 h. G-C$_3$N$_4$ was prepared by mixing the GO with melamine at a mass ratio of 1:10 and the mixture was annealed in Ar at 950 °C for 3 h.

The samples used for XPS were under continuous CV scans for 200 cycles from +0.2 to −1.0 V (vs. Ag|AgCl), in which only reduction of oxygen happened and no electrolysis of water occurred. The resultant samples were washed by Milli-Q water and separated for three times by centrifugation at 14,800 rpm to remove the alkaline electrolyte (no potassium hydroxide (KOH) was detectable by XPS). The samples were drop-casted on gold plates for XPS. The XPS analyses were conducted in an ultrahigh vacuum chamber (~10$^{-10}$ mbar) of the undulator soft x-ray spectroscopy beamline at the Australian Synchrotron, Victoria, Australia. The photonelectrons were collected by a high-resolution and high-sensitivity hemispherical electron analyzer with nine channel electron multipliers (SPECS Phoibos 150). In the scans, the excitation energy was 700 eV for better signal-to-noise ratios and the E-pass was set to 5 eV for optimum energy resolution. The excitation photon energies were calibrated by the photon energy measured on a reference Au sample and the binding energies were normalized by the C–C peak at 284.6 eV. To evaluate the signal contribution from Nafion and the gold plate, the XPS spectra of pure Nafion and a clean gold plate were also recorded. It is found that for all samples, the N and C contributions from the gold plate are negligible (less than 5% and 1%, respectively).

Transmission infrared spectra were collected using a Bruker Vertex 70 infrared spectrometer. The multilayer graphene samples before and after ORR were vacuum dried at 65 °C for 24 h and then mixed with potassium bromide (KBr, Sigma-Aldrich) by grind before pressed to pellets. Total 32 scans were collected for both the reference (pure KBr) and the samples. The reference spectrum was then subtracted from the sample spectrum to reduce the interference from the atmosphere and the equipment. The resolution is 4 cm$^{-1}$ and the scan range is from 600 to 4,000 cm$^{-1}$.

All electrochemical measurements were performed under identical conditions (the same catalyst mass loading). Taking G-NH$_3$·H$_2$O electrode as an example, the catalyst was first ultrasonically dispersed in Milli-Q water. Aqueous catalyst solution dispersion of 20 μL (2.0 mg/mL, with 0.5 wt% Nafion$^®$ solution) was then transferred onto a glassy carbon electrode (GC, 0.196 cm$^2$, Pine Research Instrumentation, USA) *via* a controlled drop casting approach then dried in air for 12 h served as a working electrode. The reference electrode was an Ag|AgCl in saturated AgCl-KCl solution and the counter electrode was a platinum wire. A flow of O$_2$ was maintained over the electrolyte (0.1 M KOH) during electrochemical measurements for continued O$_2$ saturation. Cyclic voltammograms (CVs), linear sweep voltammograms (LSVs) and rotating disk electrode (RDE) tests were carried out using a glassy carbon rotating disk electrode. The scan rate of CVs was kept at 100 mV/s while that for LSVs and RDE tests was both 10 mV/s. The data were recorded using a CHI 760D potentiostat (CH Instruments Inc., USA).

ASSOCIATED CONTENT

**Supporting Information**

Solid-state NMR, calculation of catalytic performance and subtraction of Nafion's contribution to the XPS spectrum. This material is available free of charge *via* the Internet at http://pubs.acs.org.

AUTHOR INFORMATION

**Corresponding Author**

*E-mail: luhua.li@deakin.edu.au

**Notes**

The authors declare no competing financial interest.

ACKNOWLEDGMENT

T Xing and LH Li thank the valuable discussion with Dr. Yan Jiao from Adelaide University. LH Li thanks CRGS from Deakin University for financial support. Part of the research was undertaken on the soft x-ray beamline at the Australian Synchrotron, Victoria, Australia.

Reference

1. Gong, K.; Du, F.; Xia, Z.; Durstock, M.; Dai, L. Nitrogen-Doped Carbon Nanotube Arrays with High Electrocatalytic Activity for Oxygen Reduction. *Science* **2009**, *323*, 760-764.

2. Liang, J.; Zheng, Y.; Chen, J.; Liu, J.; Hulicova-Jurcakova, D.; Jaroniec, M.; Qiao, S. Z. Facile Oxygen Reduction on a Three-Dimensionally Ordered Macroporous Graphitic $C_3N_4$/Carbon Composite Electrocatalyst. *Angew. Chem. Int. Ed.* **2012**, *51*, 3892-3896.

3. Wang, H.; Maiyalagan, T.; Wang, X. Review on Recent Progress in Nitrogen-Doped Graphene: Synthesis, Characterization, and Its Potential Applications. *ACS Catal.* **2012**, *2*, 781-794.

4. Geng, D. S.; Chen, Y.; Chen, Y. G.; Li, Y. L.; Li, R. Y.; Sun, X. L.; Ye, S. Y.; Knights, S. High Oxygen-Reduction Activity and Durability of Nitrogen-Doped Graphene. *Energy Environ. Sci.* **2011**, *4*, 760-764.

5. Li, H.; Kang, W.; Wang, L.; Yue, Q.; Xu, S.; Wang, H.; Liu, J. Synthesis of Three-Dimensional Flowerlike Nitrogen-Doped Carbons by a Copyrolysis Route and the Effect of Nitrogen Species on the Electrocatalytic Activity in Oxygen Reduction Reaction. *Carbon* **2012,** *54*, 249-257.

6. Qu, L.; Liu, Y.; Baek, J.-B.; Dai, L. Nitrogen-Doped Graphene as Efficient Metal-Free Electrocatalyst for Oxygen Reduction in Fuel Cells. *ACS Nano* **2010,** *4*, 1321-1326.

7. Rao, C. V.; Cabrera, C. R.; Ishikawa, Y. In Search of the Active Site in Nitrogen-Doped Carbon Nanotube Electrodes for the Oxygen Reduction Reaction. *J. Phys. Chem. Lett.* **2010,** *1*, 2622-2627.

8. Sheng, Z.-H.; Shao, L.; Chen, J.-J.; Bao, W.-J.; Wang, F.-B.; Xia, X.-H. Catalyst-Free Synthesis of Nitrogen-Doped Graphene Via Thermal Annealing Graphite Oxide with Melamine and Its Excellent Electrocatalysis. *ACS Nano* **2011,** *5*, 4350-4358.

9. Zhang, L.; Xia, Z. Mechanisms of Oxygen Reduction Reaction on Nitrogen-Doped Graphene for Fuel Cells. *J. Phys. Chem. C* **2011,** *115*, 11170-11176.

10. Li, Y.; Zhao, Y.; Cheng, H.; Hu, Y.; Shi, G.; Dai, L.; Qu, L. Nitrogen-Doped Graphene Quantum Dots with Oxygen-Rich Functional Groups. *J. Am. Chem. Soc.* **2011,** *134*, 15-18.

11. Kim, H.; Lee, K.; Woo, S. I.; Jung, Y. On the Mechanism of Enhanced Oxygen Reduction Reaction in Nitrogen-Doped Graphene Nanoribbons. *Phys. Chem. Chem. Phys.* **2011,** *13*, 17505-17510.

12. Liu, R.; Wu, D.; Feng, X.; Müllen, K. Nitrogen-Doped Ordered Mesoporous Graphitic Arrays with High Electrocatalytic Activity for Oxygen Reduction. *Angew. Chem. Int. Ed.* **2010,** *122*, 2619-2623.

13. Niwa, H.; Horiba, K.; Harada, Y.; Oshima, M.; Ikeda, T.; Terakura, K.; Ozaki, J.-i.; Miyata, S. X-Ray Absorption Analysis of Nitrogen Contribution to Oxygen Reduction

Reaction in Carbon Alloy Cathode Catalysts for Polymer Electrolyte Fuel Cells. *J. Power Sources* **2009,** *187*, 93-97.

14. Sharifi, T.; Hu, G.; Jia, X.; Wågberg, T. Formation of Active Sites for Oxygen Reduction Reactions by Transformation of Nitrogen Functionalities in Nitrogen-Doped Carbon Nanotubes. *ACS Nano* **2012,** *6*, 8904-8912.

15. Zheng, B.; Wang, J.; Wang, F.-B.; Xia, X.-H. Synthesis of Nitrogen Doped Graphene with High Electrocatalytic Activity toward Oxygen Reduction Reaction. *Electrochem. Commun.* **2012,** *28*, 24-26.

16. Ikeda, T.; Boero, M.; Huang, S.-F.; Terakura, K.; Oshima, M.; Ozaki, J.-i. Carbon Alloy Catalysts: Active Sites for Oxygen Reduction Reaction. *J. Phys. Chem. C* **2008,** *112*, 14706-14709.

17. Parvez, K.; Yang, S. B.; Hernandez, Y.; Winter, A.; Turchanin, A.; Feng, X. L.; Mullen, K. Nitrogen-Doped Graphene and Its Iron-Based Composite as Efficient Electrocatalysts for Oxygen Reduction Reaction. *ACS Nano* **2012,** *6*, 9541-9550.

18. Lai, L.; Potts, J. R.; Zhan, D.; Wang, L.; Poh, C. K.; Tang, C.; Gong, H.; Shen, Z.; Lin, J.; Ruoff, R. S. Exploration of the Active Center Structure of Nitrogen-Doped Graphene-Based Catalysts for Oxygen Reduction Reaction. *Energy Environ. Sci.* **2012,** *5*, 7936-7942.

19. Zhang, C.; Hao, R.; Liao, H.; Hou, Y. Synthesis of Amino-Functionalized Graphene as Metal-Free Catalyst and Exploration of the Roles of Various Nitrogen States in Oxygen Reduction Reaction. *Nano Energy* **2012,** *2*, 88-97.

20. Hammond, J. S.; Winograd, N. XPS Spectroscopic Study of Potentiostatic and Galvanostatic Oxidation of Pt Electrodes in $H_2SO_4$ and $HClO_4$. *J. Electroanal. Chem.* **1977,** *78*, 55-69.

21. Wakisaka, M.; Suzuki, H.; Mitsui, S.; Uchida, H.; Watanabe, M. Increased Oxygen Coverage at Pt-Fe Alloy Cathode for the Enhanced Oxygen Reduction Reaction Studied by EC-XPS. *J. Phys. Chem. C* **2008,** *112*, 2750-2755.

22. Pels, J. R.; Kapteijn, F.; Moulijn, J. A.; Zhu, Q.; Thomas, K. M. Evolution of Nitrogen Functionalities in Carbonaceous Materials During Pyrolysis. *Carbon* **1995,** *33*, 1641–1653.

23. Xiao, B.; Boudou, J.-P.; Thomas, K. Reactions of Nitrogen and Oxygen Surface Groups in Nanoporous Carbons under Inert and Reducing Atmospheres. *Langmuir* **2005,** *21*, 3400-3409.

24. Ganguly, A.; Sharma, S.; Papakonstantinou, P.; Hamilton, J. Probing the Thermal Deoxygenation of Graphene Oxide Using High-Resolution in situ X-Ray-Based Spectroscopies. *J. Phys. Chem. C* **2011,** *115*, 17009-17019.

25. Yang, D.; Velamakanni, A.; Bozoklu, G.; Park, S.; Stoller, M.; Piner, R. D.; Stankovich, S.; Jung, I.; Field, D. A.; Ventrice Jr, C. A. Chemical Analysis of Graphene Oxide Films after Heat and Chemical Treatments by X-Ray Photoelectron and Micro-Raman Spectroscopy. *Carbon* **2009,** *47*, 145-152.

26. Kelemen, S.; Gorbaty, M.; Kwiatek, P. Quantification of Nitrogen Forms in Argonne Premium Coals. *Energy & Fuels* **1994,** *8*, 896-906.

27. Friebel, J.; Köpsel, R. The Fate of Nitrogen During Pyrolysis of German Low Rank Coals—a Parameter Study. *Fuel* **1999,** *78*, 923-932.

28. Oh, J.; Lee, J. H.; Koo, J. C.; Choi, H. R.; Lee, Y.; Kim, T.; Luong, N. D.; Nam, J. D. Graphene Oxide Porous Paper from Amine-Functionalized Poly(Glycidyl Methacrylate)/Graphene Oxide Core-Shell Microspheres. *J. Mater. Chem.* **2010,** *20*, 9200-9204.

29. Wate, P. S.; Banerjee, S. S.; Jalota-Badhwar, A.; Mascarenhas, R. R.; Zope, K. R.; Khandare, J.; Misra, R. D. K. Cellular Imaging Using Biocompatible Dendrimer-Functionalized Graphene Oxide-Based Fluorescent Probe Anchored with Magnetic Nanoparticles. *Nanotechnology* **2012,** *23*, 415101.

30. He, H. Y.; Riedl, T.; Lerf, A.; Klinowski, J. Solid-State NMR Studies of the Structure of Graphite Oxide. *J. Phys. Chem.* **1996,** *100*, 19954-19958.

31. Lerf, A.; He, H. Y.; Riedl, T.; Forster, M.; Klinowski, J. $^{13}$C and $^{1}$H MAS NMR Studies of Graphite Oxide and Its Chemically Modified Derivatives. *Solid State Ionics* **1997,** *101*, 857-862.

32. Lin, Z.; Song, M.-K.; Ding, Y.; Liu, Y.; Liu, M.; Wong, C.-P. Facile Preparation of Nitrogen-Doped Graphene as a Metal-Free Catalyst for Oxygen Reduction Reaction. *Phys. Chem. Chem. Phys.* **2012,** *14*, 3381-3387.

33. Yu, L.; Pan, X.; Cao, X.; Hu, P.; Bao, X. Oxygen Reduction Reaction Mechanism on Nitrogen-Doped Graphene: A Density Functional Theory Study. *J. Catal.* **2011,** *282*, 183-190.

34. Boukhvalov, D. W.; Son, Y.-W. Oxygen Reduction Reactions on Pure and Nitrogen-Doped Graphene: A First-Principles Modeling. *Nanoscale* **2012,** *4*, 417-420.

35. Subramanian, N. P.; Li, X. G.; Nallathambi, V.; Kumaraguru, S. P.; Colon-Mercado, H.; Wu, G.; Lee, J. W.; Popov, B. N. Nitrogen-Modified Carbon-Based Catalysts for Oxygen Reduction Reaction in Polymer Electrolyte Membrane Fuel Cells. *J. Power Sources* **2009,** *188*, 38-44.

36. Saidi, W. A. Oxygen Reduction Electrocatalysis Using N-Doped Graphene Quantum-Dots. *J. Phys. Chem. Lett.* **2013,** *4*, 4160-4165.

37. Zhang, L.; Niu, J.; Dai, L.; Xia, Z. Effect of Microstructure of Nitrogen-Doped Graphene on Oxygen Reduction Activity in Fuel Cells. *Langmuir* **2012,** *28*, 7542-7550.

38. Huang, S.-F.; Terakura, K.; Ozaki, T.; Ikeda, T.; Boero, M.; Oshima, M.; Ozaki, J.-I.; Miyata, S. First-Principles Calculation of the Electronic Properties of Graphene Clusters

Doped with Nitrogen and Boron: Analysis of Catalytic Activity for the Oxygen Reduction Reaction. *Phys. Rev. B* **2009,** *80*, 235410.

39. Hummers, W. S.; Offeman, R. E. Preparation of Graphitic Oxide. *J. Am. Chem. Soc.* **1958,** *80*, 1339-1339.

40. Li, D.; Muller, M. B.; Gilje, S.; Kaner, R. B.; Wallace, G. G. Processable Aqueous Dispersions of Graphene Nanosheets. *Nat. Nanotechnol.* **2008,** *3*, 101-105.

TOC

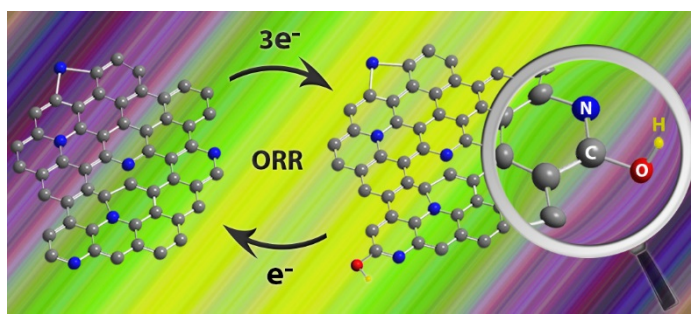